# The BCS-like gap in superconductor SmFeAsO$_{0.85}$F$_{0.15}$

T. Y. Chen,[1] Z. Tesanovic,[1] R. H. Liu,[2] X. H. Chen,[2] and C. L. Chien[1]

[1]*Department of Physics and Astronomy, Johns Hopkins University, Baltimore, MD 21218 USA*

[2]*Hefei National Laboratory for Physical Science at Microscale and Department of Physics, University of Science and Technology of China, Hefei, Anhui 230026, China*

**Since the discovery of superconductivity in the cuprates two decades ago, it has been firmly established that the CuO$_2$ plane is consequential for high $T_C$ superconductivity and a host of other very unusual properties. A new family of superconductors with the general composition of LaFeAs(O$_{1-x}$F$_x$) has recently been discovered[1-8] but with the conspicuous lacking of the CuO$_2$ planes, thus raising the tantalizing questions of the different pairing mechanisms in these oxypnictide superconductors. Intimately related to pairing in a superconductor are the superconducting gap, its value, structure, and temperature dependence. Here we report the observation of a single gap in the superconductor SmFeAs(O$_{0.85}$F$_{0.15}$) with $T_C$ = 42 K as measured by Andreev spectroscopy. The gap value of $2\Delta$ = 13.34 ±0.3 meV gives $2\Delta/k_B T_C$ = 3.68, close to the BCS prediction of 3.53. The gap decreases with temperature and vanishes at $T_C$ in a manner consistent with the Bardeen-Cooper-Schrieffer (BCS) prediction but dramatically different from that of the pseudogap behavior in the cuprate superconductors. Our results clearly indicate a nodeless gap order parameter, which is nearly isotropic in size across different sections of the Fermi surface, and are not compatible with models involving**

**antiferromagnetic fluctuations, strong correlations, t-J model, and the like, originally designed for cuprates.**

The LaFeAs($O_{1-x}F_x$) superconductors have a tetragonal structure with lattice parameters of $a \approx 4.03$ Å and $c \approx 8.73$ Å, consisting of alternating layers of quasi two-dimensional puckered LaO and FeAs planes along the $c$-axis. In the non-superconducting parent compound of LaFeAsO the Fe magnetic moments exhibit spin-density-wave antiferromagnetic ordering below about 150 K[9]. By replacing O with F, the FeAs layers can be doped with electrons resulting in the decrease of $T_N$ and the emergence of superconductivity at a doping level of x ≈ 0.15 – 0.2 in LaFeAs($O_{1-x}F_x$)[7].

An essential physical quantity in any superconductor is the superconducting gap $2\Delta$, whose value and structure are intimately related to the pairing mechanism. Band structure calculations have revealed disconnected Fermi surfaces in LaFeAs($O_{1-x}F_x$)[10], thus raising the prospect of two superconducting gaps, a situation previously encountered in $MgB_2$ [11]. In this work, we report the measurements of the superconducting gap of SmFeAs($O_{0.85}F_{0.15}$) by Andreev spectroscopy, one of the relatively few methods with which the superconducting gap can be determined. The fabrication and its superconducting properties with $T_C \approx 42$ K of the polycrystalline SmFeAs($O_{0.85}F_{0.15}$) sample have been described elsewhere[3]. We have independently confirmed the value of $T_C$ by resistance and Meissner effect measurements. For the Andreev spectroscopy measurements, we used Au tips in contact with a sample of SmFeAs($O_{0.85}F_{0.15}$) of about 5 mm in size. The measurements of differential conductance $dI/dV(V) = G(V)$ were carried out by varying the bias voltage $V$ across the contact in a variable temperature cryostat, mostly without but some with a magnetic field up to 9 T. Negative *V* is defined when electrons are injected from the tip into the superconductor.

At a normal (N) metal/superconductor (S) interface, the injected current at a bias voltage within the gap must first be converted into a supercurrent consisting of Cooper pairs of electrons with opposite spins. This can be accomplished by having the injected electron from one spin band accompanied by another electron from the opposite spin band. This is the well-known Andreev reflection process, in which it is equivalent to reflecting a hole back into the metal thus doubling the conductance within the superconducting gap. Andreev spectroscopy provides a sensitive and quantitative measurement of the gap structure of superconductors[11, 12]. The measured conductance *G(V)* can be quantitatively analyzed using the well developed theoretical models such as the modified Blonder–Tinkham–Klapwijk (BTK) model[12-14], which include effects due to thermal broadening and the less than ideal contacts as often encountered.

We first illustrate the Andreev spectroscopy measurements at 4.52 K of two well-known superconductors. The normalized conductance $G(V)/G_n$, where $G_n$ is the normal state conductance, of Nb (Fig. 1a) exhibits two peaks indicating a single gap with a value of $2\Delta = 2.84$ meV, whereas that of MgB$_2$ displays two gaps with the values of $\Delta_S = 2.75$ meV and $\Delta_L = 6.75$ meV (Fig. 1b). These results show that the Andreev spectroscopy can accurately determine the superconducting gap, or even multiple gaps only a few meV apart using polycrystalline samples of Nb and MgB$_2$.

One representative conductance of a Au/SmFeAs(O$_{0.85}$F$_{0.15}$) contact at 4.52 K within the bias voltage range of ±30 mV is shown in Figure 1c. There are two peaks with a separation of about 13.2 mV indicating a single gap, whose apparent value is close to $2\Delta = 13.6$ meV obtained from the BTK analysis, which gives an excellent description of the conductance results within ± 30 mV as shown by the red curve through the data points. However, to ensure confidence in this gap value for SmFeAs($O_{0.85}F_{0.15}$), it is essential to examine the conductance results of all the contacts and address the unexpected features. We have made conductance measurement at 4.52 K of about 15 contacts extending to ± 100 mV, with and without a magnetic field. Some of these results are shown in Figure 2, arranged in decreasing value of the contact resistance $R$, which scales inversely with the contact size. The actual values of the contact size can in principle be obtained using the Wexler formula[15] with the knowledge of the electrical resistivity, which for polycrystalline SmFeAs($O_{0.85}F_{0.15}$) pellet samples unfortunately remains quite tenuous. All the measured conductance $G(V)$ show a pronounced symmetric structure within ± 25 mV, while some contain additional small but complex structures. Despite all the extra structures, however, it is important to note that *all* the conductance curves contain the *same* feature of two peaks with an apparent separation of about 13.2 mV as indicated by the vertical dashed lines. It is therefore conclusively established that $2\Delta$ of the SmFeAs($O_{0.85}F_{0.15}$) sample is set by this value. The average of the best-fit values from all the analyzed Andreev spectra at 4.52 K is $2\Delta = 13.34 \pm 0.3$ meV. Together with $T_C = 42$ K, we obtain $2\Delta/k_B T_C = 3.68$, which is close to, although slightly higher than, the well-known universal value of $2\Delta/k_B T_C = 3.53$ within the BCS theory. Our method of identifying the superconducting gap is further corroborated by its temperature dependence, as described below.

It is often observed in the $G(V)/G_n$ of N/S contacts that a spike appears at $V = 0$, known as the zero bias anomaly (ZBA). The ZBA can have such a high intensity (>2) that it overwhelms the entire conductance spectrum. We have observed ZBA and our data show the systematic emergence of the ZBA, as illustrated in Figure 2. As the contact size becomes larger (hence smaller contact resistance $R$), the ZBA increases in intensity. At $R = 42 \, \Omega$ (Fig. 2e), the ZBA has protruded the gap structure. At $R = 4.3 \, \Omega$ (Fig. 2f), the ZBA is so large exceeding 7 (note the different scale) that it dwarfs the gap structure. The ZBA has been observed in cuprate superconductors with $d$-wave pairing[16] as well as low $T_C$ $s$-wave superconductors such as Al[17] and Nb[18]. Thus the ZBA is not exclusive to $d$-wave pairing. It is clear, however, that the ZBA is intimately related to the superconductivity, since ZBA does not appear at temperature above $T_C$, nor in contacts of normal metals. The ZBA is also related to the contact size, hence the contact resistance, as illustrated in Figure 2. Consequently, the conductance results containing a dominant ZBA contribution (e.g., Fig. 2f) would render the extraction of the gap structure fruitless. We have found that a 9 T magnetic field at 4.52 K has very little effect on both the extraneous structure and the superconducting gap feature, as shown in Figure 2(d, f), in sharp contrast to the claim that the ZBA can be substantially reduced by a magnetic field of only a few teslas[19].

In addition to the main gap, all the conductance curves also show extra features, some extending to ± 100 mV. However, while the actual causes are not known, these features appear related to superconductivity since they disappear at higher temperatures. It is worth mentioning that the conductance curves of high quality single-crystal cuprate oxide superconductors such as $La_{2-x}Sr_xCuO4$, also show rather complex additional features outside the gap structure[20]. Another unusual feature shown in Figure 2 is the asymmetric background, always with $G(-V) > G(V)$. However, the asymmetrical background remains at temperatures much higher than $T_C$, hence its origin appears unrelated to superconductivity but is perhaps due to the large mismatch of conductivity between metal and oxide conductors. One notes that most STM results of cuprate superconductors display an acute asymmetrical background[21]. In contrast, contacts with low $T_C$ metallic superconductors, such as Nb and $MgB_2$, do not exhibit these features. The conductance curves shown in Figure 1(a, b) are flat extending to very large voltages and show the expected symmetry of $G(V) = G(-V)$.

The temperature dependence of the superconducting gap, in addition to the gap value, is also of great importance. We have determined the temperature dependence of the gap of $SmFeAsO_{0.85}F_{0.15}$ by varying the temperature for one contact and measured more than 70 conductance curves at various temperatures from 4.5 K to 60 K. These results are summarized in Figure 3a. At low temperatures, the gap structure is resolved and has a high intensity. The separation of the two peaks, hence the value of the gap, decreases with increasing temperature, and evolves into a single unresolved peak with decreasing width. The intensity of the central peak also decreases with increasing temperature and vanishes at $T_C$ and remains so at temperature above $T_C$. The gap value, obtained from analyses of the conductance curves, show a temperature dependence shown in Figure 3b. The measured temperature dependence of $\Delta$ is close to the BCS prediction shown by the dashed curve.

Although we have used a polycrystalline $SmFeAs(O_{0.85}F_{0.15})$ sample, the measured result of a single gap may not be the result of averaging. Due to the small contact size (especially those with large contact resistances) the gap value is most often measured from a single grain. Since numerous contacts have yielded essentially the same gap value at a given temperature, the gap structure in SmFeAs($O_{0.85}F_{0.15}$) is likely to be nearly isotropic. Ultimately, key questions such as the possibility of anisotropic gap and the nodal gap associated with *d*- or *p*-wave pairing can only be addressed using single crystal specimens. Our results also show that if multiple gaps exist in SmFeAs($O_{0.85}F_{0.15}$), their values must be very close to each other since only a single sharply defined gap, without any discernable structure, has been observed as shown in Figure 1c.

The gap value and its temperature dependence in SmFeAs($O_{0.85}F_{0.15}$) are dramatically different from those found in the pseudogap regime of high $T_C$ cuprate superconductors. The pseudogap of the cuprate superconductors is large, much larger than the value from the BCS prediction, and most importantly, it is essentially temperature *independent*[21]. Furthermore, not only does the tunneling pseudogap persist at the same value, it can even be resolved at temperatures much higher than $T_C$, as high as $3T_C$. Even in some Andreev reflection experiments, where the pseudogap gap disappears closer to $T_C$, its $T = 0$ value is far in excess of BCS values for $2\Delta/k_BT_C = 3.53$ (*s*-wave) and $2\Delta/k_BT_C = 4.28$ (*d*-wave). These unique characteristics of the pseudogap in the cuprate superconductors are absent in SmFeAs($O_{0.85}F_{0.15}$).

The gap value $\Delta$ of the SmFeAs($O_{0.85}F_{0.15}$) superconductor and its temperature dependence are close to the BCS predictions. However, the exceptionally high $T_C$ of 40 - 55 K seems to have exceeded the limit of the electron-phonon coupling. The actual mechanism of superconductivity in SmFeAs($O_{0.85}F_{0.15}$) has yet to be determined, and there are already a number of theoretical proposals. Our results provide clear evidence of two central aspects of the pairing mechanism in the superconducting SmFeAs($O_{0.85}F_{0.15}$): it is a single-gap superconductor, its gap exhibiting a BCS-type temperature dependence, and there are no signatures of the pseudogap structure as featured prominently in the cuprate superconductors. Furthermore, we do not see evidence of nodal structure in the gap and the observed $2\Delta/k_B T_C$ is notably less than the weak-coupling d- and p-wave values. Thus, our results imply a nodeless, BCS-type gap and set strong constraints on the ongoing theoretical research.

Figure Captions

**Figure 1 | Representative Andreev spectra at 4.52 K**. **a**, A Au/Nb contact shows one gap. **b,** A Co/MgB$_2$ contact shows two distinct gaps. c, A Au/SmFeAs(O$_{0.85}$F$_{0.15}$) contact shows one gap. The open circles are the experimental data and the solid curves in **a** and **c** are the best-fit results to the modified BTK model [14] with parameters shown. Our methods of experiment and data analysis are detailed in the Supplementary Information.

**Figure 2 | Andreev spectra of Au/SmFeAs(O$_{0.85}$F$_{0.15}$) point contacts at 4.52 K with various contact resistance. a-f**, Spectra are arranged in decreasing contact resistances $R$, where open circles are the experimental data and solid curves are the best fit to the modified BTK model with the parameters and contact resistance listed in each figure. The vertical dashed lines are at ±6.6 meV to indicate the common features. Blue curve in **d** and **f** were results taken in an external magnetic field of $H = 9$ T.

**Figure 3 | Temperature dependence of the gap**. **a**, Andreev spectra from 4.51K to 56.8K of a Au/SmFeAs(O$_{0.85}$F$_{0.15}$) contact taken approximately every 0.8 K showing the decrease of the gap value and the eventual disappearance of the gap structure at $T_C \approx 42$ K. **b,** Temperature dependence of the gap value (open circles) obtained from the modified BTK fit of SmFeAs(O$_{0.85}$F$_{0.15}$) is close to the BCS theory prediction (dotted curve). The error bar at 4.51K shows ±1 s.e.m from the average of 15 spectra for different contacts. The inset shows the resistive transition at $T_C$.

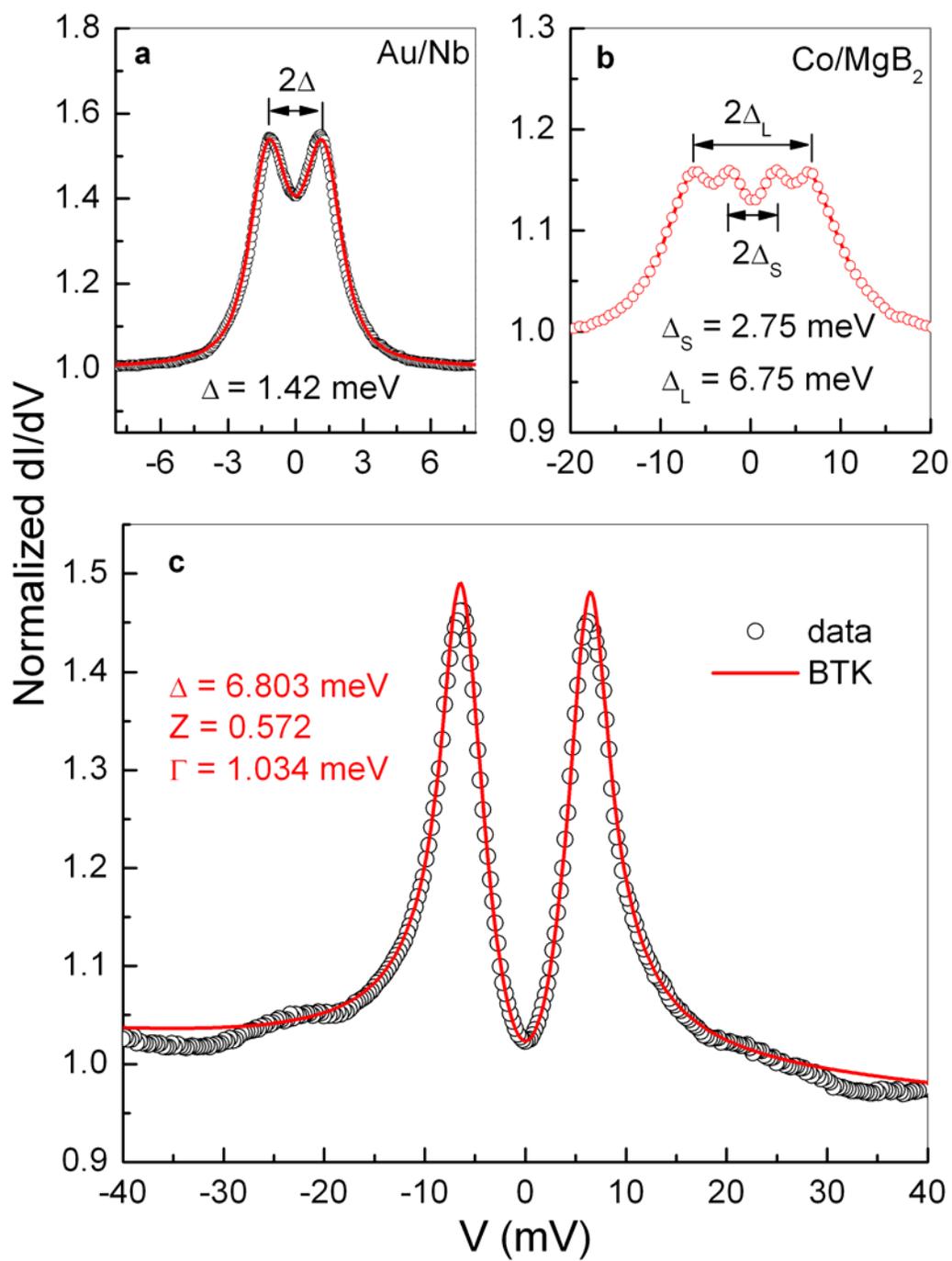

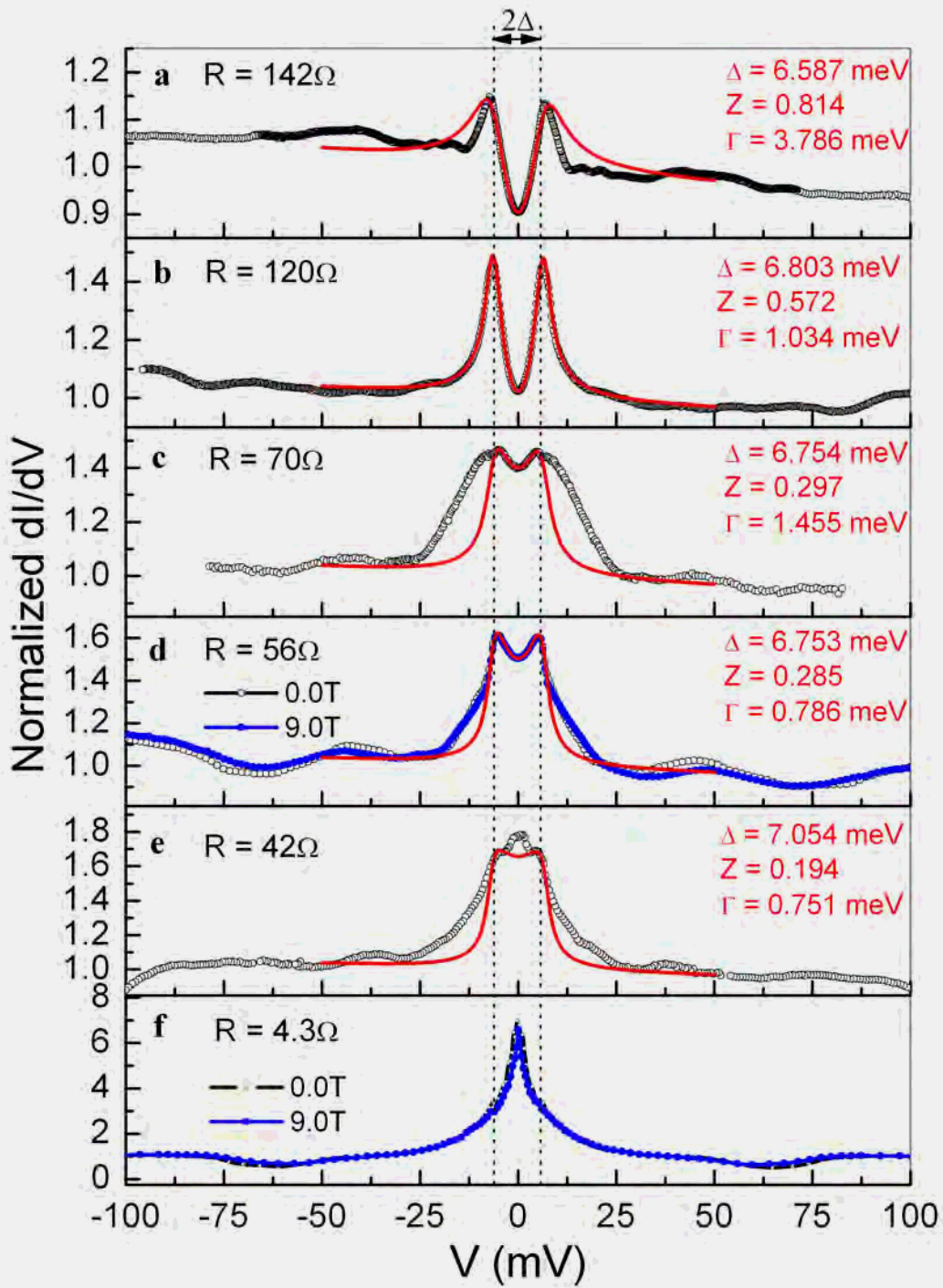

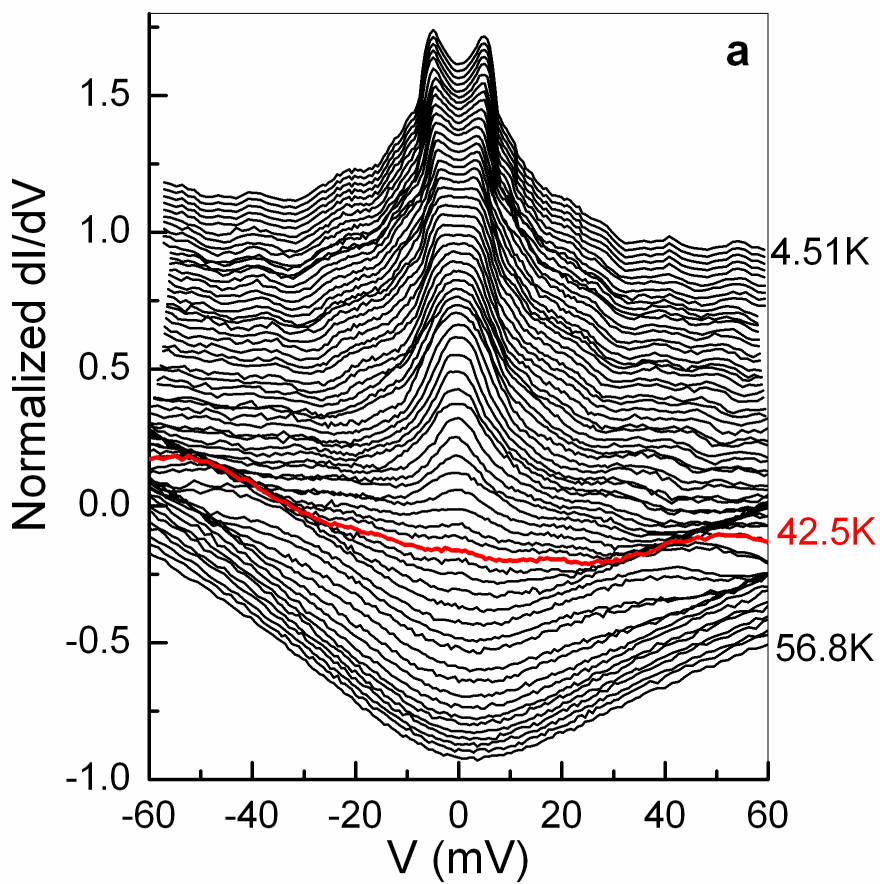

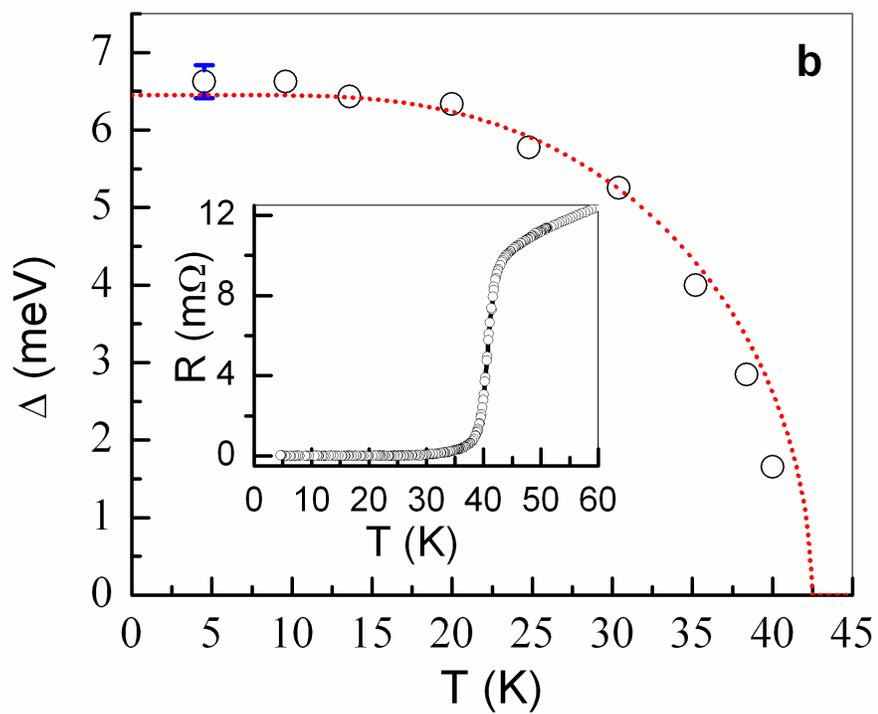

Supplementary Information

The current $I_{NS}$ as a function of the bias voltage $V$ through a N/S point contact interface can be described by the BTK theory [13] in which the interfacial barrier is represented by a δ function with a dimensionless strength factor $Z$:

$$I_{NS}(V) = C\int_{-\infty}^{\infty}[f(E-eV) - f(E)][1 + A(E) - B(E)]dE,$$

where $A(E)$ and $B(E)$ are the probability of Andreev reflection and normal reflection and can be calculated from the coherence factors $u_0$ and $v_0$ described below, $f(E)$ is the Fermi distribution function, and $C$ is a constant depending on the contact area, density of states and Fermi velocity. In reality, the quarsiparticle lifetime is shortened by inelastic scattering near the N/S interface, leading to a smearing effect on the conductance. The inelastic effects can be included in the BTK theory by introducing an imaginary component $\Gamma$ into the energy thus the coherence factors are [14]

$$\tilde{u}_0^2 = \frac{1}{2}\left[1 + \frac{\sqrt{(E+i\Gamma)^2 - \Delta^2}}{E+i\Gamma}\right]$$

$$\tilde{v}_0^2 = \frac{1}{2}\left[1 - \frac{\sqrt{(E+i\Gamma)^2 - \Delta^2}}{E+i\Gamma}\right],$$

where $\Delta$ is the superconducting gap. The probability $A(E)$ and $B(E)$ are calculated as follows,

$$A(E) = a \cdot a^*$$
$$B(E) = b \cdot b^*$$
$$a = \tilde{u}_0 \tilde{v}_0 / \gamma$$
$$b = -(\tilde{u}_0^2 - \tilde{v}_0^2)(Z^2 + iZ)/\gamma$$
$$\gamma = \tilde{u}_0^2 + (\tilde{u}_0^2 - \tilde{v}_0^2)Z^2$$

When $\Gamma$ is zero, this model becomes the original BTK model. It is worthwhile to note that the model turns into the Dynes's model [22] accounting for the inelastic effects for tunneling when $Z \rightarrow \infty$, which is often utilized in analyzing spectra of scanning tunneling microscopy for superconductors.

Supplementary Figure 1 illustrates the effect of $Z$ and $\Gamma$ on the normalized conductance curve $G(V)/G_n$ with the method outlined above. Four conductance curves have been calculated for $Z = 0$, 0.25, 0.5, and 1.0, with the other parameters fixed at $T = 0$, $\Delta = 6.5$ meV, and $\Gamma = 0$, as shown in Supplementary Figure 1**a**. At $Z = 0$, the conductance is a bell-shape curve in which $G(V)/G_n = 2$ for $|V|<\Delta$ and $G(V)/G_n = 1$ for $|V|>>\Delta$. As $Z$ increases, the Andreev reflection at low voltages is suppressed, and peaks appear at $\pm\Delta$. The effect of $\Gamma$, however, is very different. Andreev reflection is suppressed at low $V$ but also smeared around $\pm\Delta$ when $\Gamma$ increases, as shown in Supplementary Figure 1**b** where four conductance curves for $\Gamma/\Delta=0$, 0.1, 0.2, and 0.5 are calculated with the other parameters fixed at $T = 0$, $\Delta = 6.5$ meV, and $Z = 0$. When effect of both $Z$ and $\Gamma$ are included, and at finite temperatures, two conductance peaks generally appear in $G(V)/G_n$ around $\pm\Delta$, as illustrated by some calculated examples in Supplementary Figure 1**c-d** with the parameters used listed in the figure.

The Andreev spectroscopy was performed using a normal metal tip in contact with the superconductor. Our experimental setup is schematically shown in Supplementary Figure 2**a**. Two electrodes are attached to the tip and the sample. The tip and the sample were enclosed in a vacuum jacket inserted in a cryostat with controllable temperature and magnetic field up to 9 T. The measurements of *I* and *dI/dV* versus *V* curves were made using a four-probe method and a lock-in technique. Because of the asymmetric background, $G_n$ cannot be determined by simply increasing *V*. Supplementary Figure 2**b** illustrates the process of determining $G_n$. We linearly fit the data between 25 mV <|*V*| < 50 mV and choose the point at *V*=0 as $G_n$. Then the normalized data is analyzed using the above model with an extra linear background.

Some representative data at different temperatures for Figure 3 in the manuscript are shown in Supplementary Figure 3. The open circles are experimental data and solid curves are best-fit using the modified BTK model. One notes the fine vertical scale of the plots, showing that the data can be well described by the model even at 40K when the conductance change is only a few percent. All the conductance curves share a similar background.

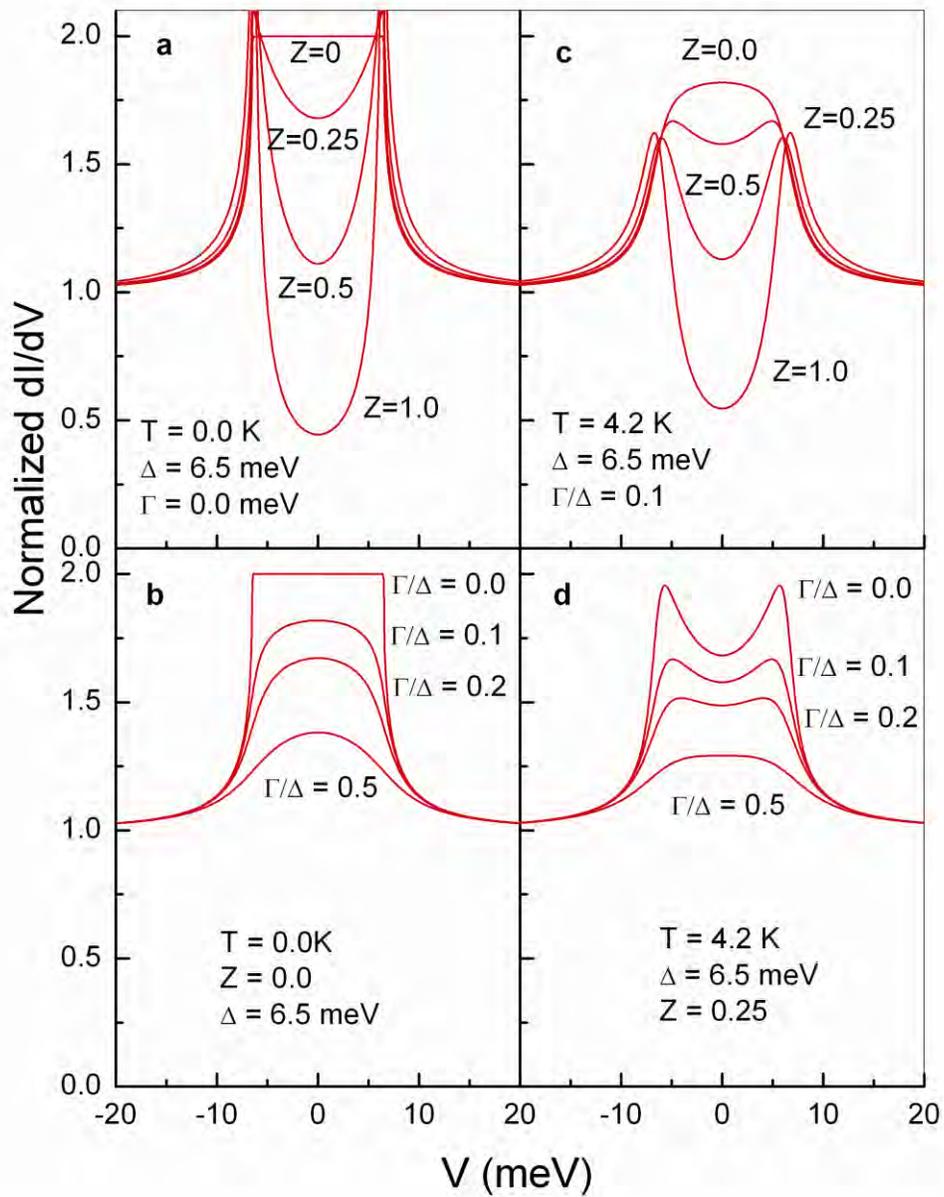

Supplementary Figure 1 | Calculated conductance curves using the modified BTK model: **a** for $Z = 0$, 0.25, 0.5, 1.0 with $T$, $\Delta$ and $\Gamma$ fixed at 0, 6.5 meV, and 0, **b** for $\Gamma/\Delta =$ 0, 0.1, 0.2, 0.5 with $T$, $\Delta$ and $Z$ fixed at 0, 6.5 meV, and 0, **c** for $Z = 0$, 0.25, 0.5, 1.0 with $T$, $\Delta$ and $\Gamma/\Delta$ fixed at 4.2 K, 6.5 meV, and 0.1, and **d** for $\Gamma/\Delta =$ 0, 0.1, 0.2, 0.5 with $T$, $\Delta$ and $Z$ fixed at 4.2 K, 6.5 meV, and 0.25.

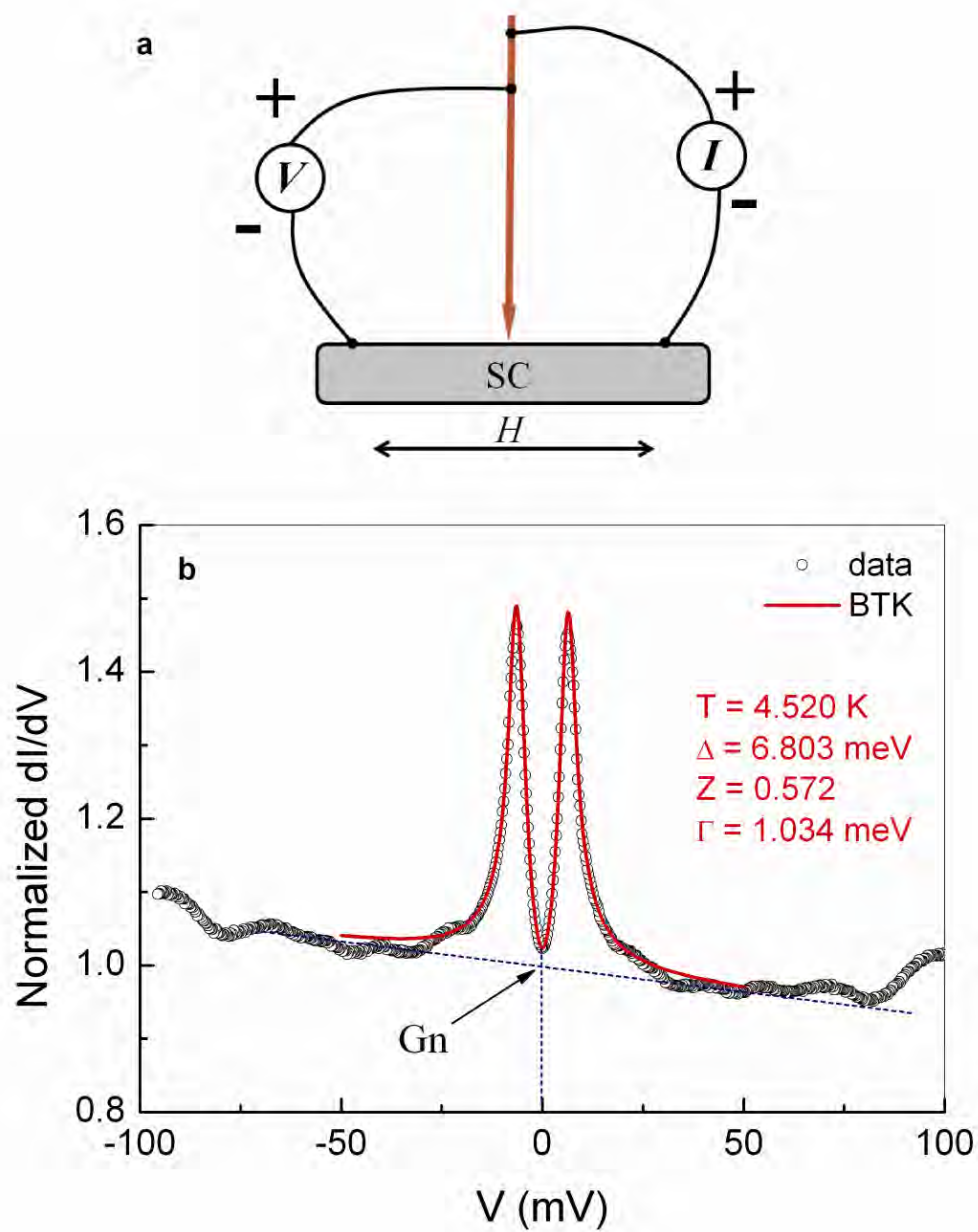

Supplementary Figure 2 | **a** Schematics of experimental setup of a gold tip in contact with a superconductor (SC) with the electrical leads attached, and **b** determination of the normalization conductance $G_n$ (Open circles are the data and solid line is the best fit to the modified BTK model with parameters listed in the figure).

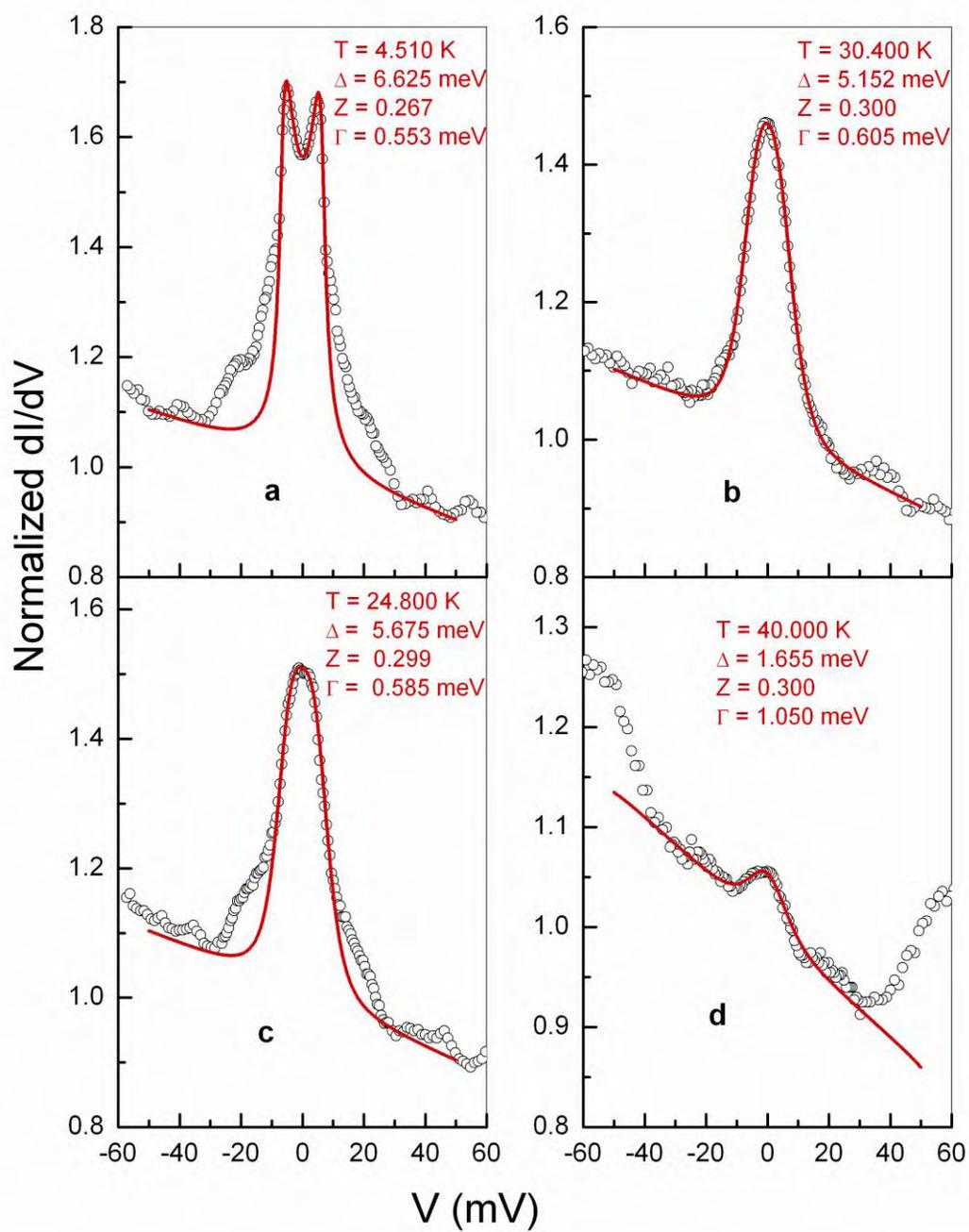

Supplementary Figure 3 | Representative best fits to the data at different temperatures with the parameters listed in each panel.